\documentclass[11pt]{article}
	
	\newcommand{\blind}{0}
    \makeatletter
    \renewcommand\section{\@startsection {section}{1}{\z@}%
                                       {-3.5ex \@plus -1ex \@minus -.2ex}%
                                       {2.3ex \@plus.2ex}%
                                       {\normalfont\fontfamily{phv}\fontsize{16}{19}\bfseries}}
    \renewcommand\subsection{\@startsection{subsection}{2}{\z@}%
                                         {-3.25ex\@plus -1ex \@minus -.2ex}%
                                         {1.5ex \@plus .2ex}%
                                         {\normalfont\fontfamily{phv}\fontsize{14}{17}\bfseries}}
    \renewcommand\subsubsection{\@startsection{subsubsection}{3}{\z@}%
                                        {-3.25ex\@plus -1ex \@minus -.2ex}%
                                         {1.5ex \@plus .2ex}%
                                         {\normalfont\normalsize\fontfamily{phv}\fontsize{14}{17}\selectfont}}
    \renewcommand{\l@section}[2]{\@dottedtocline{1}{0em}{2.3em}{#1}{}}
    \renewcommand{\l@subsection}[2]{\@dottedtocline{2}{1.5em}{3.2em}{#1}{}}
    \renewcommand{\l@subsubsection}[2]{\@dottedtocline{3}{3.8em}{4.1em}{#1}{}}                      
    \makeatother

	\usepackage{amsmath}
	\usepackage{graphicx}
	\usepackage{enumerate}
	\usepackage{xcolor}
	\usepackage{url} 
    \usepackage[colorinlistoftodos]{todonotes} 
    \usepackage{longtable}
    \usepackage{lineno}
    \usepackage{floatrow}
    \usepackage{float}
    \usepackage{caption}
    \usepackage{booktabs}
    \usepackage{multirow}
    \usepackage{xcolor}
    \usepackage{hyperref}
    \usepackage[style=apa]{biblatex}
    \usepackage{xr}
    \usepackage{enumitem}
    \usepackage[T1]{fontenc}
    \usepackage[final]{changes}
    \usepackage{pdfpages} 

\begin{document}
		
	\def\spacingset#1{\renewcommand{\baselinestretch}%
		{#1}\small\normalsize} \spacingset{1}
		
    \if0\blind
    {
        \title{\bf Analyzing recreational fishing effort - Gender differences and the impact of Covid-19}
        \author{Julia S. Schmid\thanks{Corresponding author (email: jschmid@ualberta.ca)} $^1$, Sean Simmons$^2$, Mark S. Poesch$^3$, Pouria Ramazi$^6$, Mark A. Lewis$^{1,3,4,5}$ \\
        \small
           $^1$Department of Mathematical and Statistical Sciences, University of Alberta, Edmonton, Alberta, Canada \\
         \small
         $^2$Angler’s Atlas, Goldstream Publishing, Prince George, British Columbia, Canada \\
         \small
         $^3$Department of Biological Sciences, University of Alberta, Edmonton, Alberta, Canada \\
         \small
         $^4$Department of Biology, University of Victoria, Victoria, British Columbia, Canada \\
         \small
         $^5$Department of Mathematics and Statistics, University of Victoria, Victoria, British Columbia, Canada \\
         \small
         $^6$Department of Mathematics and Statistics, Brock University, St. Catharines, Ontario, Canada}
        \date{}
        \maketitle
    } \fi
		
	\if1\blind
	   {
        \title{\bf \emph{Analyzing recreational fishing effort - Gender differences and the impact of Covid-19}}
		\author{Author information is purposely removed for double-blind review}
			
    \bigskip
	\bigskip
	\bigskip
	\begin{center}
		{\LARGE\bf \emph{Analyzing recreational fishing effort - Gender differences and the impact of Covid-19}}
		\end{center}
		\medskip
		} \fi
	\bigskip
\newpage		
\spacingset{2}
\begin{abstract}
    \replaced{Recreational fishing}{Fishing} is a\replaced{n important economic driver and provides multiple social benefits}{valuable recreational activity in our society}. 
    To \replaced{predict}{assess future} fishing activity, identifying variables related to \replaced{variation}{differences in fishing activity}, such as gender or Covid-19, is helpful. 
    We conducted a Canada-wide email survey of users of an online fishing platform and analyzed responses \replaced{focusing}{with a focus} on gender, the impact of Covid-19, and variables directly related to \replaced{fishing}{fisher} effort.
    Genders (90\deleted{.1}\% \replaced{men}{male} and \replaced{10}{9.9}\% \replaced{women}{female respondents}) significantly differed in demographics, socioeconomic status, and fishing skills but \replaced{showed}{were} similar \deleted{in }fishing preferences, \replaced{fishing}{fisher} effort in terms of trip frequency, and travel distance. 
    \replaced{Covid-19 altered trip frequency for almost half of fishers, with changes varying by gender and activity level.}
    {For almost half of the fishers, Covid-19 caused a change in trip frequency, determined by the activity level and gender of the fisher.} 
    A Bayesian network revealed \deleted{that} travel distance \replaced{as}{was} the main determinant of trip frequency\replaced{,}{ and} negatively \replaced{impacting}{impacted the} fishing activity \replaced{for}{of} 61\% of\deleted{the} fishers\added{, with fishing expertise also playing a role}. 
    \deleted{Fisher effort was also directly related to fishing expertise.}
    \replaced{The results suggest that among active fishers, socio-economic differences between genders do not drive fishing effort, but responses to Covid-19 were gender-specific. 
    Recognizing these patterns is critical for equitable policy-making and accurate socio-ecological models, thereby improving resource management and sustainability.}{ 
    The study shows how online surveys and Bayesian networks can help understand the relationship between fishers' characteristics and activity and predict future fishing trends.}
\end{abstract}

\noindent%

\newpage

\section{Introduction} \label{s:intro}

Recreational fishing \deleted{has a high value in our society as it} is an important economic driver (\cite{DFO2019}) and provides multiple social benefits such as stress release and connection to nature (\cite{arlinghaus2009recreational, young2016fishers, DFO2019, floyd2006social}).
Freshwater recreational fisheries are complex adaptive systems, not only defined by ecological feedbacks but also by social drivers (\cite{arlinghaus2017understanding}). 
Hence, effective sustainable management requires an understanding of both the ecological environment such as fish stock dynamics, and the social environment such as fishing behavior. 
Identifying how variables, such as demographics and socioeconomic status, are related to fishing activity can help to provide insights into the system and to improve fishing sustainability and fisher satisfaction \deleted{in the future }(\cite{arlinghaus2009recreational, barcellini2013recreational, brownscombe2019future, Sbragaglia2023Preparing}).

Participation in recreational fishing activities varies among people with different demographics. 
A long-term study in Texas revealed that women and young people go fishing less often and less consistently than older people and men (\cite{fedler2001dropping}). 
\replaced{Most recent data in Canada showed men represented 79\% of fishers with ages}{In 2015, approximately 79\% of Canadian fishers were male and the fisher's age distribution was} skewed to older age groups (\cite{DFO2019}). 
Reasons for variable fishing participation could be differing fishing motivations and \deleted{levels of} satisfaction \added{levels} between the genders. 
Socioeconomic backgrounds could also play a role because recreational fishing is a leisure activity (\cite{schroeder2006he, mostegl2011catch, hickley1998recreational}).
Most previous studies focused on variables influencing the participation of males and females in fishing activities. 
However, they disregarded how the \replaced{fishing}{fisher} effort of participating men and women was influenced by different factors.

\replaced{Fishing}{Fisher} effort of participating fishers, such as the frequency of fishing trips, can also  be influenced by demographics and socio-economic status.
In addition, the distance and accessibility of water bodies, weather conditions and the global Covid-19 pandemic could \replaced{alter}{lead to variations} in \replaced{fishing}{fisher} effort\added{, e.g., by reducing fisher access to fishing sites} (\cite{birdsong2021recreational, gundelund2022investigating, britton2023global}). 
Uncovering variables that \replaced{impact fishing}{lead to spatial and temporal variation in fisher} effort can help to enable participatory management and \replaced{forecasting}{the prediction of future} fishing activity trends (\cite{kleiber2015gender, hickley1998recreational}). 

Surveys \replaced{were often}{are} used to obtain data related to fishing participation and \replaced{fishing}{fisher} effort (\cite{rees2017socio, czarkowski2021socio, arlinghaus2003socio, hinrichs2021motivations, ditton2004substitutability}). 
\replaced{Often these survey data}{Relationships between different variables in the survey data} were \deleted{commonly} analyzed using traditional statistical methods such as regression models, analysis of variance (ANOVA) and chi-square tests (\cite{rees2017socio, czarkowski2021socio, cinner2006socioeconomic}).
These methods can be restrictive, for instance, in terms of handling missing data that could result from missing answers to individual survey questions, only analyzing interactions between pairs of factors disregarding the system as a whole (e.g., chi-square tests), 
or assuming linearity or certain distributional forms for the relationships between the variables (e.g., ANOVA).

Bayesian networks overcome these restrictions and can outperform traditional statistical methods \deleted{such as logistic regression models }(\cite{Scanagatta2019, Song2022}). 
\replaced{Bayesian networks use c}{C}onditional dependencies between different variables \deleted{are} visualized in a network that can be learned from data and used to detect non-linear relationships between variables (\cite{Scanagatta2019, Ramazi2021MEE}).
Several studies have combined survey data with Bayesian networks in different fields for different purposes such as supporting medical decisions (\cite{constantinou2016complex}), analyzing illegal crossing behavior of pedestrians (\cite{ma2020analysis}), 
predicting mountain pine beetle infestations in forests (\cite{Ramazi2021MEE, ramazi2021predicting}) and assessing the risk of safety levels in university laboratories (\cite{zhao2023risk}).

In this study, gender differences among fishers and conditional dependencies between possible variables affecting \replaced{fishing}{fisher} effort in terms of fishing frequency were analyzed.
More specifically, three research questions were addressed: 
1. Do \replaced{women and men who fish recreationally}{male and female recreational fishers} differ in terms of demographics, socioeconomic status, fishing preferences, characteristics and effort?
2. How did Covid-19 affect the \replaced{fishing}{fisher} effort of fishers of different genders and \added{of} different levels of fishing activity?
3. What variables can be used to \replaced{predict}{determine} \replaced{fishing}{fisher} effort?    
An online survey across Canada was conducted with questions related to fishers' demographics and socioeconomic status, and fishing characteristics, motivations, preferences and effort.
Relationships between the variables were examined both statistically and via a Bayesian network. 
The latter made it possible to analyze direct and indirect probabilistic dependencies between the variables and to extract main determinants.
\deleted{The results give insights into the role of gender and differing influences of variables on fisher effort which can be used to 
detect potential fishing hotspots in time and space.}

\section{Materials and Methods} \label{s:methods}

\subsection{\emph{Email survey}} \label{s:methods.1}
\subsubsection{\emph{Survey respondents}} \label{s:methods.11}
    An email survey was sent on July 26, 2023, to all subscribers (126,431) to the newsletter of the Angler's Atlas platform, a leading online platform for fishers in Canada (\url{www.anglersatlas.com}). 
    Survey participants could win one of four gift cards of CA\$100 (1 in 333 chance to win). 
    Until August 31, 2023, 44,589 recipients opened the email (35\%), 2,287 recipients clicked on the link of the survey and 1,689 recipients responded which resulted in a response rate of 3.8\% of the recipients that opened the email. 
    The study was reviewed and approved by the Research Ethics Board of the Alberta Research Information Services (ARISE, University of Alberta), study ID \texttt{MS5\_Pro00102610}.
    
\subsubsection{\emph{Survey questions}} \label{s:methods.12}
    The survey comprised 36 questions of seven categories: (1) ten questions on demographics and socioeconomic status, (2) four on fishing characteristics, (3) four on fisher behavior, (4) eight questions on the preferable water-body environment, (5) two on the preferable weather conditions, (6) four on the usage of the Angler's Atlas platform, (7) two on the impact of Covid-19, and two additional open questions to share additional information (see appendix). 
    Questions were in the format of multiple-choice, single-select and open-text responses.
    \added{See Supplementary Information for the actual questions.}

\subsection{\emph{Data preprocessing}} \label{s:methods.2}
    Non-relevant responses from fishers that indicated no fishing region or a\deleted{nother} fishing region \added{other} than Canada were removed (n=5). 
    Respondents who indicated \added{as gender} ``female" were considered women, and respondents who indicated ``male" were considered men. 
    \deleted{Both expressions are used in the manuscirpt and refer to the gender.}
    One percent of the respondents chose not to specify their gender or selected “Other” (n=16) and were \replaced{therefore}{therfore} not included in the study. 
    See SI methods for answers under “Other”.
    
    Text responses were individually assessed and classified into the specified categories of each question. Postal codes of origin were assigned to the respective province. 
    Continuous variables (short-distance trip frequency, long-distance trip frequency, household size, and fishing experience) were binned into four levels \replaced{with equal frequency}{based on the quantiles of the data}.     
    
    For questions with categorical responses (``Fishing reason" and ``Use of Angler's Atlas") binary dummy variables were created for each response option (seven variables for ``Fishing reason" and six variables for ``Use of Angler's Atlas"). 
    See Table \ref{tab:Covariates} for an overview of all variables and their levels.
    
    Of the 1,668 responses (samples), 703 were incomplete, meaning that at least for one variable the value was not available (NA) (Fig. \ref{fig:NAN_values}). 
    NA values appeared either because the participant provided no answer, the participant chose the option not to answer (e.g., for ``Income") or the answer of the participant did not fit into any reasonable predefined category. 

\subsection{\emph{Data evaluation}} \label{s:methods.3}
\subsubsection{\emph{Differences between genders}} \label{s:methods.31}
    Gender and age distributions were compared to two surveys among fishing licence purchasers to evaluate how representative the survey respondents were. 
    In 2015, Fisheries and Oceans Canada (DFO) surveyed Canada-wide active fishers using a stratified, systematic random sampling method and gained a participation rate of 4.4\% (115,372 of 2,639,224 active residential and non-residential fishers, \cite{DFO2019}).
    Moreover, data on 270,120 fishing licence purchasers fishing in the province of British Columbia in the year 2022 were \replaced{available}{provided by the Freshwater Fisheries Society of BC}.

    The chi-square test of independence was applied to detect differences between the frequencies of responses from \replaced{women and men}{females and males} \added{using the \texttt{chi2\_contingency()} function of the \texttt{scipy.stats} module in \texttt{Python}}. 
    NA values were excluded.

\subsubsection{\emph{Factors related to fishing effort}} \label{s:methods.32}
    The preprocessed data set was further adjusted for learning Bayesian networks. 
    Samples with more than four NA values were removed (n=68). 
    The variables ``Province of residence'' and ``Main fishing province'' were grouped into five geographical categories (see Table \ref{tab:Covariates} and SI Methods for details).
    The preprocessed data set included 1600 samples of 46 variables in seven categories (Table \ref{tab:Covariates}).
    
    A Bayesian network was used to represent the set of variables and their probabilistic dependencies (\cite{koller2009probabilistic}). It consisted of a directed acyclic graph and conditional probability distributions that were presented as tables in the case of discrete variables. 
    The construction of the Bayesian network based on data was divided into two steps: (i) structure learning, the estimation of the network structure that capture\replaced{s}{d} the dependencies between the variables (the edges in the network), and (ii) parameter learning, the estimation of the \deleted{(conditional)} probability distributions of \replaced{each}{individual} variable\deleted{s} \added{conditioned on its parents }(\cite{Ramazi2021MEE}). 
    The Bayesian network was learned with the \texttt{bnlearn} package (version 4.9.1) in \texttt{R} (version 4.3.2) (\cite{scutari2009learning, Hansen2023}).
 
    For (i), learning the structure based on an incomplete data set, the structural expectation-maximization algorithm was used to find the ``best" network structure in an iterative process, consisting of repeated expectation and maximization steps (\texttt{structural.em()} in \texttt{bnlearn}, \cite{Friedman1997learning, hernandez2013learning, Scanagatta2019}). 
    In the first expectation step, the NA values in the incomplete data set were imputed using an initial empty network structure to obtain a complete data set. In the maximization step, the complete data set was used in the score-based greedy hill-climbing to find the network structure with the maximum Bayesian Information Criteria (BIC) score in 1,000 iteration steps (\cite{beretta2018learning, gamez2011learning}). 
    The new network structure was then used in the following expectation step to impute the NA values in the incomplete data set again. 
    The stopping criteria for the expectation-maximization algorithm was set to five iterations. 
    The resulting network structure with the highest score was chosen as the ``best" network structure of the Bayesian network.

    For (ii), learning the parameters of the Bayesian network, maximum likelihood estimation was used with the incomplete data set (\texttt{bn.fit()} in \texttt{bnlearn}). 
    Given the best network structure, the conditional probabilities of each node was obtained based on the relative frequencies of variable values of locally complete samples (i.e., samples that had values for the node and its parent nodes).
    Conditional probabilities were expressed as relative frequencies in the results to be consistent with the first part of the analysis \added{in which differences between genders were analyzed}. 

    The strengths of dependencies between the variables were measured by generating 2,500 additional Bayesian network structures from bootstrap samples (\texttt{boot.strength()} in \texttt{bnlearn}, \cite{broom2012model}). 
    Bootstrap samples were randomly sampled from the original data set with replacement to obtain a new data set of the same size (\cite{Friedman2013, imoto2002}). 
    The strengths of the dependencies were estimated by their relative frequencies in the network structures   
    and can be interpreted as probabilities for the inclusion of the edge in the Bayesian network. 

    See Supplementary Information for an R script illustrating the methods described above.

\section{Results} \label{s:results}

\subsection{\emph{Survey respondents}} \label{s:results.1}
    Of the 1,668 considered survey respondents across Canada, 90.1\% were \replaced{men}{male} and 9.9\% were \replaced{women}{female}. 
    Email survey respondents represented around 0.06\% of the Canadian active fishers in the year 2015 (about 2,831,000) and comprised a smaller fraction of \replaced{women}{female fishers} (21\% \replaced{women}{females}, \cite{DFO2019}). 
    Moreover, email survey respondents showed an older age distribution as compared to active fishers in the DFO survey of 2015, with 46\% in the age group 46-65 years compared to 42\% (45-64 years), 26\% in the age group 26-45 years compared to 37\% (25-44 years) and 22\% in the age group 66 years and older compared to 14\% (65 years and older) (\cite{DFO2019}, Fig. \ref{fig:Age_Distribution}A). 
    Like in the DFO survey, \replaced{women}{females} were on average younger than \replaced{men}{males} (45 years and 49 years in the DFO survey, \cite{DFO2019}). 
    
    In the province of British Columbia, a higher fraction of email survey respondents than fishing license purchasers in 2021 were male (87\% of 480 respondents vs. 77\% of 270,120 purchasers, Fig. \ref{fig:Age_Distribution}B). 
    \replaced{E}{Moreover, e}mail survey \replaced{respondents}{responents} were on average younger than fishing licence purchasers (\(\chi^2\) p < 0.001)(Fig. \ref{fig:Age_Distribution}B).    
    
\subsection{\emph{Gender differences}} \label{s:results.2}
    Between the genders, significant differences were found in variables of all the considered categories.

\begin{figure}[H]
    \includegraphics[width=1\linewidth]{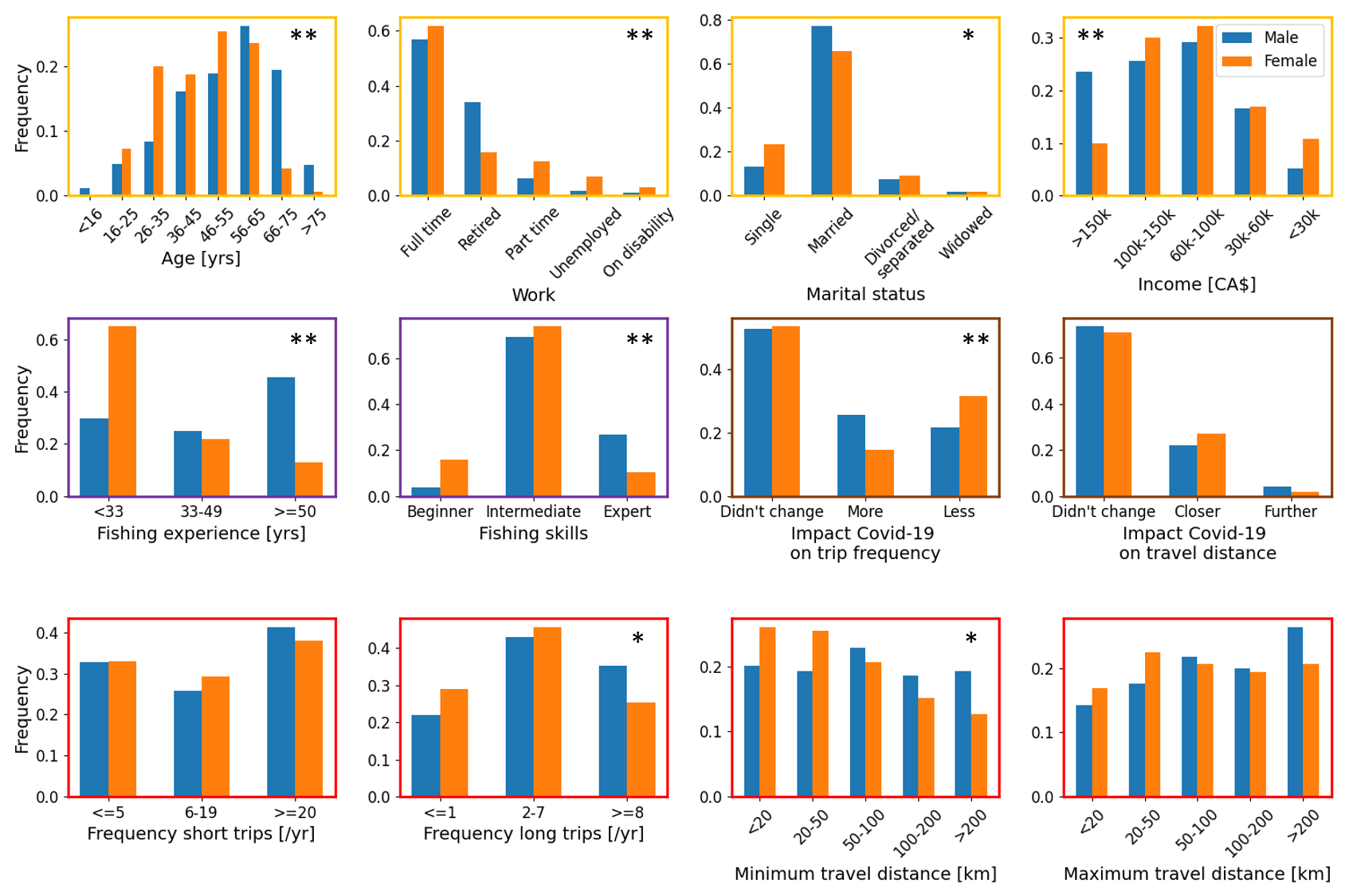}
    \caption{Responses of male (blue) and female (orange) fishers. Colors of frames indicate different categories of the variables. \(\chi^2\) test for difference in proportions of males and females, *: p < 0.05, **: p < 0.01. See SI for remaining variables.}
    \label{fig:Covariate_Distribution_main}
\end{figure}

    Regarding demographics and socioeconomic status, the distributions of the variables Age and Work differed (\(\chi^2\) p < 0.01, Table \ref{tab:chi-square-results}). 
    The majority of \deleted{the} \replaced{men}{male participants} were in the age group 56-65 years, whereas \replaced{women}{female participants} were mainly in the age group 46-55 years (Fig. \ref{fig:Covariate_Distribution_main}). 
    Most participants were employed full-time (57\%), but more \replaced{men}{males} were retired compared to \replaced{women}{females} (34\% vs. 16\%), whereas more \replaced{women}{females} were part-time employed or unemployed (19\% vs. 8\%, Fig. \ref{fig:Covariate_Distribution_main}). 
    Differences occurred also in the variables Marital status (p < 0.05), Income (p < 0.01) and Access to boat (p < 0.01, Table \ref{tab:chi-square-results}). 
    \replaced{Women}{Females} had a higher proportion of the status single or divorced/separated (32\% vs. 9\%) and more \replaced{men}{male respondents} earned more than \$150,000 per year (24\% vs. 10\%, Fig. \ref{fig:Covariate_Distribution_main}). 
    \replaced{Men}{Male fishers} were more likely to have access to a boat (79\% vs. 66\%, Fig. \ref{fig:Covariate_Distribution_appendix}). 
    Independent of gender, most participants had a college, technical training or university degree (71\%), a household size of two or more persons (92\%), were married (76\%) and earned between \$60,000 and \$150,000 per year (55\%, Figs. \ref{fig:Covariate_Distribution_main}, \ref{fig:Covariate_Distribution_appendix}). 
    Most fishers had always access to a vehicle (97\%) and resided in the provinces of British Columbia, Alberta or Ontario (87\%, Fig. \ref{fig:Covariate_Distribution_appendix}).

    Fishing characteristics differed significantly between \replaced{women and men}{males and females} regarding the variables Fishing experience and Fishing skills (\(\chi^2\) p < 0.01) (Fig. \ref{fig:Covariate_Distribution_appendix}). 
    The Bayesian network revealed that in each level of Fishing experience, \replaced{men}{male fishers} dominated (Fig. \ref{fig:CPDs_appendix}A). 
    45\% of the \replaced{men}{males} reported having more than 50 years of fishing experience (vs. 13\% of \replaced{women}{females}, Fig. \ref{fig:Covariate_Distribution_main}). 
    The proportion of \replaced{women}{female fishers} increased from 3\% in high Fishing experience (more than 50 years) to 18\% in low Fishing experience (less than 33 years, Fig. \ref{fig:CPDs_appendix}A). 
    Fishing skills were mostly intermediate regardless of gender (70\%), but a higher fraction of \replaced{women}{females} reported being beginners compared to \replaced{men}{males} (16\% vs. 4\%), who reported more often being experts (69\% vs. 10\%, Fig. \ref{fig:Covariate_Distribution_main}). 
    Responses to the reasons for fishing were similar for both genders except that a higher fraction of \replaced{women}{females} indicated the reasons ``being outside'' (\(\chi^2\) p < 0.01) and ``food'' (p < 0.05), and a higher fraction of \replaced{men}{males} indicated ``sport'' as a reason (p < 0.01, Fig. \ref{fig:Covariate_Distribution_appendix}. 
    In general, reasons for fishing were relaxation (74\%), enjoyment (85\%) and being outside (66\%) rather than fishing for food (34\%), sport (35\%) or competition (7\%, Fig. \ref{fig:Covariate_Distribution_appendix}). 
    Most participants cited British Columbia, Alberta or Ontario as their main provinces for fishing trips (85\%, Fig. \ref{fig:Covariate_Distribution_appendix}).

    Fisher behavior in terms of the frequencies of short-distance trips and the maximum travel distance did not significantly differ between \replaced{women and men}{males and females} (Fig. \ref{fig:Covariate_Distribution_main}). 
    \replaced{Men}{Male respondents} had more long-distance trips (\(\chi^2\) p < 0.05) and \replaced{women}{females} tended to have shorter minimum travel distances (p < 0.05). 
    Most respondents indicated to have more than 20 short-distance trips (41\%) and two to seven long-distance trips per year (43\%, Fig. \ref{fig:Covariate_Distribution_main}). 

    Regarding the water-body environment, responses to \replaced{preferred}{prefered} water-body type and ocean fishing were similar for both genders. 
    Compared to \replaced{men}{males}, a higher portion of \replaced{women}{females} indicated that fishing regulations were not important (50\% vs. 41\%) or they preferred water bodies without regulations (24\% vs. 19\%, \(\chi^2\) p < 0.01, Table \ref{tab:chi-square-results}, Fig. \ref{fig:Covariate_Distribution_appendix}). 
    More \replaced{women}{females} compared to \replaced{men}{males} preferred a quiet environment (90\%, \(\chi^2\) p < 0.01) and more \replaced{men}{males} compared to \replaced{women}{females} fished only from the boat (44\%, p < 0.01)
    Independent of gender, most respondents preferred to fish in small lakes (36\%) or had no preference for any particular water-body type (27\%), and did not fish in the ocean in addition to freshwater (64\%, Fig. \ref{fig:Covariate_Distribution_appendix}). 

    In weather preferences, there were differences between the two genders regarding \replaced{preferred}{prefered} hot, windy or calm conditions (\(\chi^2\) p < 0.05, Table \ref{tab:chi-square-results}, Fig. \ref{fig:Covariate_Distribution_appendix}). 
    The majority of respondents stated that hot weather is not important (37\%) or they might cancel fishing in hot (23\%) or windy weather (42\%), 
    that rainy weather has a medium impact on their decision about going fishing (49\%), that they might or would usually fish in low air pressure (49\%), 
    that they would usually fish in calm weather (61\%) and they might (28\%) or would not care (33\%) to fish in cold weather (Fig. \ref{fig:Covariate_Distribution_appendix}).

    See SI Results for the results regarding the usage of the Angler's Atlas online platform. 
    
\subsection{\emph{Impact of Covid-19}} \label{s:results.3}
    Covid-19 had different effects on the frequency of fishing between genders (\(\chi^2\) p < 0.01, Table \ref{tab:chi-square-results}) and effects on the travel distance were similar. 
    \replaced{Women}{Female fishers} tended to go fishing more often due to Covid-19 (32\%) compared to \replaced{men}{male fishers} (26\%), while a higher fraction of \replaced{men}{males} went fishing less often (22\% \replaced{men}{males} vs. 15\% \replaced{women}{females}) (Fig. \ref{fig:Covariate_Distribution_main}). 
    Most respondents indicated that Covid-19 did not affect their travel distance (73\%, Fig. \ref{fig:Covariate_Distribution_main}).

    Additionally, the frequency of fishing trips remained the same for more than half of the fishers during Covid-19 (53\%, Fig. \ref{fig:Covariate_Distribution_main}, irrespective of how active they were (Fig. \ref{fig:CPDs_main}A). 
    If Covid-19 caused a change, fishers with lower trip frequency (less than six short-distance trips per year) went fishing more often (74\%), and fishers with a high trip frequency (more than 20 short-distance trips per year) went fishing less often (63\%). 
    
    The travel distance only changed for 27\% of fishers due to Covid-19 (Figs. \ref{fig:Covariate_Distribution_main}, \ref{fig:CPDs_main}B). 
    Fishers who went more fishing due to Covid-19 were more likely to travel to closer water bodies due to Covid-19 than fishers who did not change their behavior or went less fishing due to Covid-19 (Fig. \ref{fig:CPDs_appendix}J).

\subsection{\emph{Variables related to fishing effort}} \label{s:results.4}
\begin{figure}[H]
  \includegraphics[width=1.1\linewidth]{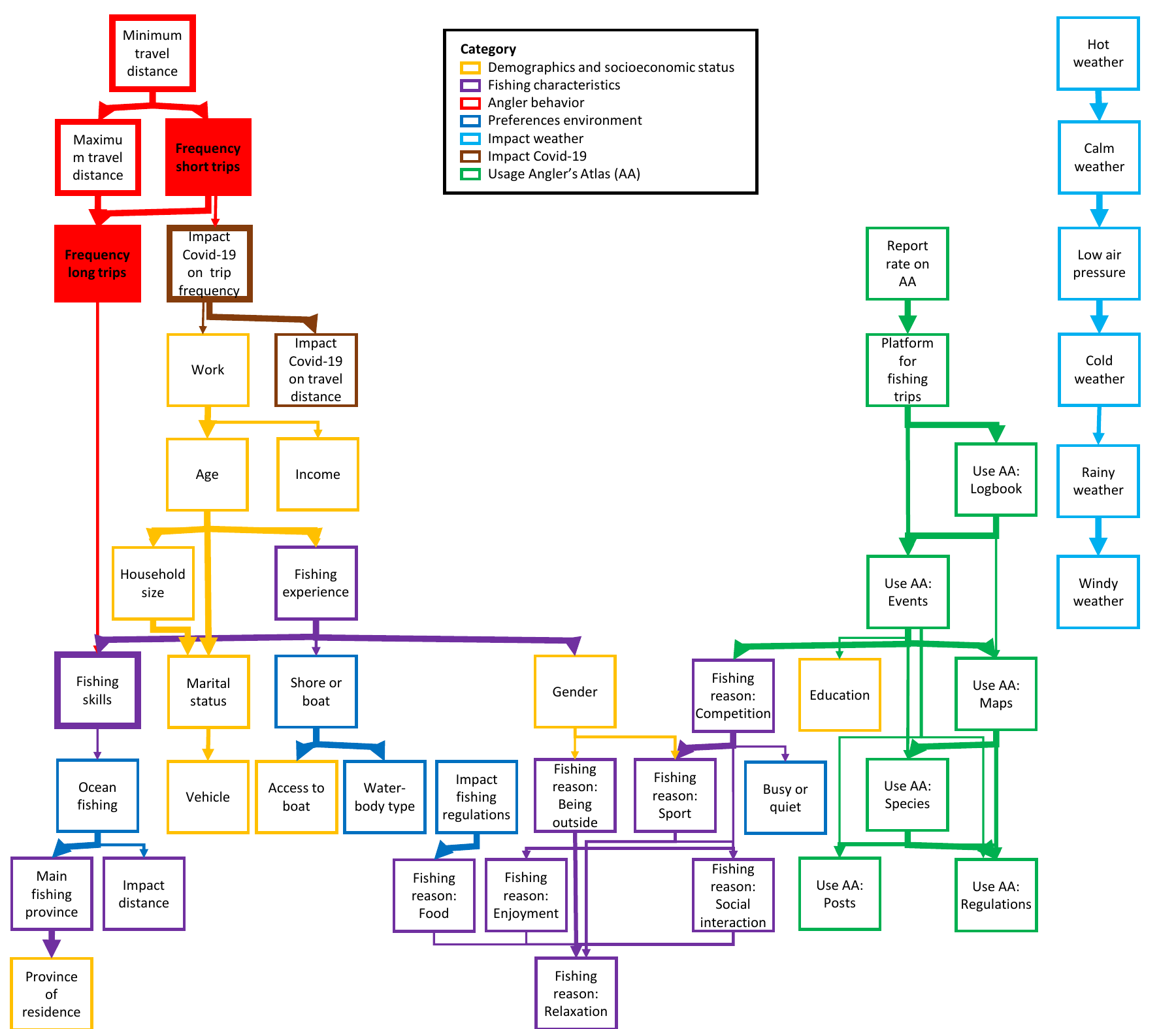}
  \caption{Dependencies between \replaced{fishing}{fisher} effort and other variables in a Bayesian network. 
  \replaced{Fishing}{Fisher} effort, i.e., the frequency of short- and long-distance trips, are the red-colored boxes and directly related variables are in a bold frame. 
  Colors of nodes refer to different categories and thicknesses of edges to different strengths of dependencies. 
  Edges indicated conditional dependencies between the nodes. Note that the direction of an edge does not necessarily indicate causality.}
  \label{fig:BN}
\end{figure}

\begin{figure}[H]
    \includegraphics[width=0.8\linewidth]{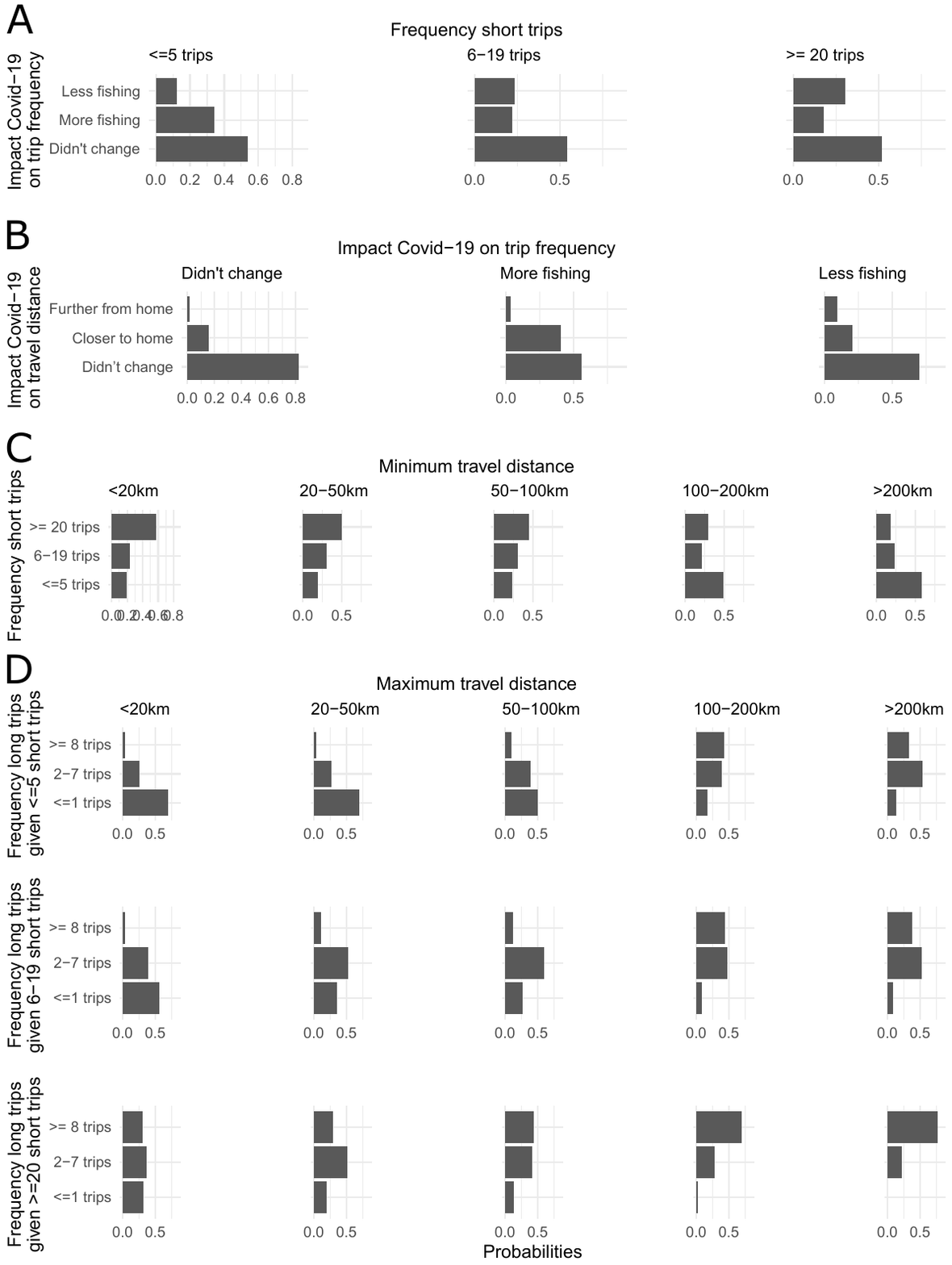}
    \caption{Relationships between variables connected to \replaced{fishing}{fisher} effort. Bars show conditional probabilities between (A) the frequency of short-distance trips and the impact of Covid-19 on the trip frequency, (B) the impact of Covid-19 on trip frequency and the impact of Covid-19 on travel disctance, 
    (C) the minimum travel distance and the frequency of short-distance trips and 
    (D) the maximum travel distance, the frequency of short trips and long trips. 
    See Fig. \ref{fig:BN} for on overview of the dependencies.}
    \label{fig:CPDs_main}
\end{figure}

    The Bayesian network enabled to detect direct and indirect relationships as well as dependenc\replaced{e}{y} strengths between the variables. 
    Direct conditional dependencies occurred in the network mostly between variables of the same category (e.g., ``Demographics and socioeconomic status" or ``Fishing characteristics", Fig. \ref{fig:BN}). 
    Strong relationships between variables were found in the category ``Fisher behavior" and in the category ``Impact weather".
    Variables on preferences regarding the water body environment (e.g., busy or quiet environment) were scattered in the network and connected to variables of different categories. 
    ``Impact weather" variables were not related to variables of other categories.

    \replaced{Fishing}{Fisher} effort was directly related to travel distance, fishing skills and the impact of Covid-19 on trip frequency.

    \textit{Relationships between fishing skills and \replaced{fishing}{fisher} effort:}
    Most fishers indicated to have intermediate fishing skills, independent from fishing experience and the long-distance trip frequency. 
    Fishers doing more than eight long-distance trips per year were likelier to indicate being fishing experts than fishers with fewer long-distance trips, irrespective of their fishing experience (Fig. \ref{fig:CPDs_appendix}H).

    \textit{Relationships between travel distance and \replaced{fishing}{fisher} effort:}
    Travel distances were directly related to the frequency of fishing trips (Fig. \ref{fig:BN}). 
    Minimum and maximum travel distance were the only variables that trip frequencies were strongly dependent on. 
    The minimum and maximum travel distance of a respondent were mostly in the same range. 
    For instance, if the minimum travel distance was less than 20 km, it was very likely that the maximum travel distance was also less than 20 km (Fig. \ref{fig:CPDs_appendix}D). 
    The shorter the minimum travel distance was, the more likely the fisher did at least 20 short-distance trips per year (<100 km distance) (Fig. \ref{fig:CPDs_main}C).
    Most fishers with at least 20 short-distance trips per year had also more than eight long-distance trips per year (>100 km distance) (Fig. \ref{fig:CPDs_main}D). 
    61\% of the fishers cared about the distance when choosing the water body (Fig. \ref{fig:Covariate_Distribution_appendix}), but the distance became less relevant if the fisher did mainly ocean fishing (Fig. \ref{fig:CPDs_appendix}G).

\section{Discussion} \label{s:discussion}

An online survey combined with Bayesian networks made it possible to identify variables directly related to fishing effort and the role of gender among participating fishers.

Most variables in fishing preferences and \replaced{fishing}{fisher} effort were similar between genders although demographics, \replaced{socioeconomic}{socioeconimic} status and fishing skills differed. 
\replaced{Fishing}{Fisher} effort, in terms of annual trip frequency, and travel distances differed in the minimum travel distance and the frequency of long-distance trips between the genders. 
Covid-19 impacted the fishing activity of almost half of the respondents, whereby female or less-active fishers tended to increase the number of short-distance trips, 
and male fishers or very active fishers tended to decrease their short-distance trip frequency. 
The main \replaced{determinant}{determint} of trip frequency was trip distance, which negatively impacted the fishing activity of more than half of the fishers. 
Besides trip distance, fishing experience and fishing skills were associated with long-distance fishing frequency.
Water-body environment and weather preferences as well as fishing reasons were only indirectly or not related to \replaced{fishing}{fisher} efforts.

\deleted{In this study, we only analyzed active fishers who overcame the barriers to participate in fishing.}
\added{The underrepresentation of women in the online survey aligns with the findings of previous studies (\cite{floyd2006social, arlinghaus2004management}),
whereby barriers such as less leisure time due to family responsibilities, limited education and technical training, traditional gender roles, and power imbalances were identified (\cite{Bradford2023, Mayer2019sharing, salmi2018invisible, fedler2001dropping, floyd2006social}).}
\added{This study specifically examined active fishers who \replaced{were able to participate despite potential challenges or restrictions}{managed to overcome obstacles to participate in fishing}.}
\replaced{Significant gender differences in fishing effort and reasons were anticipated, however, the results challenge the idea that these factors consistently limit women's engagement. 
For example, despite disparities in age, work, income, and fishing experience, the frequency of short fishing trips was similar between genders.
This indicates that initial barriers to participation in fishing, rather than persistent socio-economic or role-related constraints, are the main cause of gender differences in fishing.}
{Still, significant differences in fisher effort and fishing reasons between genders were expected, but only occured in some of the analyzed variables.}
\replaced{Further investigation into the factors that enable women to overcome these barriers, such as the presence of support networks or autonomy, is needed.}
{The low participation of female fishers in the online survey is consistent with other studies (Arlinghaus and Mehner, 2004; Floyd et al., 2006).}
\deleted{Stated arguments for lower participation of women in fishing or fishing-related activities were often less leisure time due to family duties such as childcare, 
limited education and technical training, and traditional gender roles or gendered power imbalances (Bradford et al., 2023; Fedler and Ditton, 2001;
316 Floyd et al., 2006; Mayer and Le Bourdais, 2019; Salmi and Sonck-Rautio, 2018).
The here detected differences between genders in age, work, income and fishing experience and skills together with similar fishing effort among active fishers suggest that 
other factors than socioeconomic status and fishing characteristics are responsible for the lower participation of women in fishing activities in general, 
and that fishing activity of women who go fishing is barely affected by the differences.}

\replaced{Regardless}{Irrespective} of gender, \replaced{the}{responses regarding} fishing preferences and reasons \replaced{of respondents aligned}{were in agreement} with \replaced{findings from previous}{other} studies\replaced{.}{ which indicates that the study respondents ecompassed different types of fishers.} 
\replaced{Most respondents fished for}{For instance, fishing reasons being rather} relaxation, enjoyment and \added{a connection to} nature \added{rather} than food or competition\replaced{.}{,} 
\replaced{They also preferred}{and the preference of} calm fishing environments\replaced{, which is consistent with studies from the central United States and the West Coast}
{are in line with answers of studies across the central United States and the west coast of the United States} (\cite{hinrichs2021motivations, young2019adaptation}). 
\added{These similarities suggest that our sample represents a diverse range of recreational fishers, as seen in broader studies.}

The impact of Covid-19 on \replaced{fishing}{fisher} effort was related to gender and activity level of the fisher.
The fact that Covid-19 led to a change of fishing participation and \replaced{fishing}{fisher} effort was also \replaced{reported}{found} in previous studies (\cite{howarth2021covid, midway2021covid, britton2023global, audzijonyte2023high}). 
Previous studies focused on possible reasons for changes in fishing activity such as improved mental and physical health and the listing of fishing as an essential activity for more fishing activity, 
or uncertain accessibility and the closure of national parks for reduced fishing activity due to Covid-19 (\cite{howarth2021covid, paradis2021can}), 
but they did not distinguish between fishers of different demographic backgrounds or socioeconomic status in their \replaced{analyses}{ananlyses}.

The strong connection between \replaced{fishing}{fisher} effort to the travel distance is in agreement with previous studies in which most fishers chose water bodies in close proximity for their fishing activity (\cite{camp2018angler, jalali2022angling}). 
These studies focused on reasons for specific water body choices of fishers, but did not analyze \replaced{whether and how}{how and if} the distance affected their trip frequency.
The connection between fishing experience, fishing skills and long-distance fishing frequency can be explained by the fact that highly specialized fishers would do long travel-distances trips 
for specific target fish species, catch rates and bag limits (\cite{camp2018angler,lewin2021travels, dabrowksa2017understanding}).

Responses on the impact of different weather conditions on fishing activity were not related to \replaced{fishing}{fisher} effort.
Fisher satisfaction and activity was also independent from weather conditions such as wind speed and maximum air temperature in previous studies (\cite{hunt2007predicting, gundelund2022investigating}).
Still, the email survey showed that some fishers would cancel their fishing trip at certain weather conditions.
The missing dependencies between weather and \replaced{fishing}{fisher} effort in the Bayesian network were likely due to the missing information of the actual weather conditions at fisher's water bodies of choice. 
Moreover, the network showed independence between weather preferences and preferences regarding the water body environment and fishing reasons. 
This suggests that weather conditions could provide additional useful information for predicting fisher behavior.

The data \replaced{represented}{representats} only angler behavior of a subgroup of fishers in Canada. 
\replaced{The sampling method was}{In} \replaced{convenience}{convencience} sampling, \added{meaning that} the email survey was only sent to fishers who had signed up for the Angler's Atlas platform newsletter, 
and the analysis only included responses \replaced{from}{of} fishers who responded to the email and were willing to share their information, 
which represents \replaced{approximately}{around} 0.06\% of the active fishers in Canada in the year 2015 (\cite{DFO2019, etikan2016comparison}).  
A younger age distribution of email survey participants compared to fishing licence purchasers in the province of British Columbia is consistent with a previous study (\cite{Gundelund2020}). 
The main provinces of residency of email survey respondents (British Columbia, Alberta and Ontario) were different from the DFO survey in 2015 (Ontario and Quebec, \cite{DFO2019}). 
The smaller participation of the French part of Canada could result from the fact that the online platform is only available in English, and, hence, less promoted and used in the province of Quebec. 
Moreover, only residents of Canada were considered in this study, but a substantial part of recreational fishers in Canada are non-residential (\cite{DFO2019}).
Although the identified factors influencing fisher behavior may be biased, the results provide valuable information about a subgroup of fishers that can be useful for analyzing the entire social system.
 
To validate the generality of the identified relationships, future studies can compare Bayesian networks based on data from similar surveys in different countries or different subgroups of fishers such as fishing license buyers. 
Moreover, the presented methodological approach can be extended to analyze not only \replaced{fishing}{fisher} effort but also variables related to fishing participation by applying Bayesian networks to a broader group using fishing survey data of all residents, not only active fishers.  

The Bayesian network can also be used to make predictions of \replaced{fishing}{fisher} effort in specific regions, given a set of variables from the fishers living in that region. 
For instance, knowing the number of fishers in a region and their minimum travel distances can be used to predict the number of annual short- and long-distance trips, and thus the total annual fisher pressure in the region \added{(\cite{fischer2023boosting})}. 
Other incomplete subsets of the variables in the network can be used to make these predictions. 
Such predictions can help \replaced{improve inclusivity and fishers management}{identify spatial hotspots of fisher pressure}.

\added{The findings highlight the need to incorporate gender differences in social-ecological models to avoid misrepresenting human behavior and undermining resource management. 
While fishing effort was similar among women and men, socio-economic and behavioral differences shaped their participation and responses to external factors such as Covid-19. 
Ignoring these differences risks ineffective policies and unequal access to fishing opportunities.
Recognizing these patterns is critical for equitable policy-making and accurate socio-ecological models, ensuring sustainable management while maintaining social and economic benefits.}

\section{Acknowledgements}
We acknowledge the support of the Government of Canada’s New Frontiers in Research Fund (NFRF), NFRFR-2021-00265.
We thank Adrian Clarke and Adeleida Bingham from the Freshwater Fisheries Society of BC for providing the numbers of fishing licence sales in British Columbia, Canada.
PR acknowledges funding from an NSERC Discovery Grant RGPIN-2022-05199.

\section{Data Availability Statement}
The data that support the findings of this study are available on request from the corresponding author. The data are not publicly available due to privacy or ethical restrictions.

\section{Conflict of Interest Statement}
The authors declare no conflict of interest.

\printbibliography

\newpage
\spacingset{1.2}
\begin{table}[H]
    \centering
    \scriptsize 
    \caption{Variables and their respective levels used in the Bayesian Network. BC - British Columbia, ON - Ontario, AB - Alberta, QB - Quebec, NB - New Brunswick, SK - Saskatchewan, MT - Manitoba, NS - Nova Scotia, NL - Newfoundland and Labrador, TT - The Territories, PEI - Prince Edward Island}
    \label{tab:Covariates}
    \begin{tabular}{|p{2.2cm}|p{3.3cm}|p{10.5cm}|} 
        \hline 
        Category &  Variable & Definition and value\\ 
        \cline{1-1}\cline{2-2}\cline{3-3}
        Demographics and & Age & 0: \textless16 years; 1: 16-25 years; 2: 26-35 years; 3: 36-45 years; 4: 46-55 years; 5: 56-65 years; 6: 66-75 years; 7: \textgreater75 years\\
        \cline{2-3}
        socioeconomic & Gender & 0: Male; 1: Female\\ 
        \cline{2-3}
        status & Marital status & 0: Single; 1: Married; 2: Divorced/separated; 3: Widowed\\ 
        \cline{2-3}
        & Education & 0: 12th grade or less; 1: High school graduate; 2: College graduate; 3: Some college / Technical training; 4: University degree; 5: Post graduate degree (Masters or Doctorate)\\ 
        \cline{2-3}
        & Income & 0: \textgreater\$150,000; 1: \$100,000-\$150,000; 2: \$60,000-\$100,000; 3: \$30,000-\$60,000; 4: \textless\$30,000 \\ 
        \cline{2-3}
        & Household size & 0: 1 person; 1: 2 persons; 3: \textgreater2 persons \\ 
        \cline{2-3}
        & Work & 0: Employed full time; 1: Retired; 2: Employed part time; 3: Unemployed; 4: On disability \\ 
        \cline{2-3}
        & Province of residence & 0: BC (0); 1: ON (1); 2: AB (2); 3: SK (5), MT (6), TT (9); 4: QB (3), NB (4), NS (7), NL (8), PEI (10)\\
        \cline{2-3}
        & Vehicle & 0: Yes; 1: No; 2: Sometimes\\ 
        \cline{2-2}
        & Access to boat &\\ 
        \hline 
        Fishing & Fishing experience & 0: \textless33 years; 1: 33-49 years; 2: $\ge$50 years \\ 
        \cline{2-3}
        characteristics & Fishing skills & 0: Beginner; 1: Intermediate; 2: Expert\\ 
        \cline{2-3}
        & Main fishing province & 0: BC (0); 1: ON (1); 2: AB (2); 3: SK (5), MT (6), TT (9); 4: QB (3), NB (4), NS (7), NL (8), PEI (10) \\ 
        \cline{2-3}
        & Impact of distance & 0: Yes; 1: No\\ 
        \cline{2-3}
        & Fishing reason (boolean) & Relaxation, Enjoyment, Social interaction, Food, Sport, Being outside, Competition \\ 
        \hline 
        Fisher behavior &  Frequency short trips & 0: $\le$5 trips; 1: 6-19 trips; 2: $\ge$20 trips\\
        \cline{2-3}
        & Frequency long trips & 0: $\le$1 trip; 1: 2-7 trips; 2: $\ge$8 trips\\
        \cline{2-3}
        & Minimum travel distance & 0: \textless20km; 1: 20-50km; 2: 50-100km;\\
        \cline{2-2}
        & Maximum travel distance & 3: 100-200km; 4: \textgreater200km \\
        \hline 
        Preferences & Water body type& 0: River; 1: Small lake; 2: Big lake; 3: Not important\\
        \cline{2-3}
        environment & Busy or quiet & 0: Not important, 1: Quiet; 2: Busy\\
        \cline{2-3}
        & Shore or boat & 0: Both; 1: Boat; 2: Shore\\
        \cline{2-3}
        & Impact fishing regulations & 0: Not important; 1: Prefer with fish size and bag size limitations; 2: Prefer with fish size limitations; 3: Prefer with bag size limitations; 4: Prefer with catch-and-release; 5: Prefer without regulations\\
        \cline{2-3}
        & Ocean fishing & 0: No; 1: Yes, but preferably freshwater fishing; 2: Yes, mainly in the ocean\\
        \hline
        Impact weather & Hot weather & 0: Doesn't matter;\\
        \cline{2-2}
        & Rainy weather & 1: Might cancel fishing;\\
        \cline{2-2}
        & Windy weather & 2: Might go fishing;\\
        \cline{2-2}
        & Calm weather & 3: Usually cancel fishing;\\
        \cline{2-2}
        & Cold weather & 4: Usually go fishing\\
        \cline{2-2}
        & Low air pressure & \\
        \hline
        Usage Angler's & Use AA (boolean) & Maps, Species, Regulations, Logbook, Events, Posts\\
        \cline{2-3}
        Atlas (AA) & Platform for fishing trips & 0: None; 1: App, 2: Website \\
        \cline{2-3}
        & Report rate on AA & 0: No trips at all; 1: \textless50\%; 2: \textgreater50\%; 3: All trips\\
        \hline
        Impact Covid-19 & Impact Covid-19 on trip frequency & 0: Didn't change; 1: More fishing; 2: Less fishing\\
        \cline{2-3}
        & Impact Covid-19 on travel distance & 0: Didn't change; 1: Closer to home; 2: Further from home\\
        \hline
    \end{tabular}
\end{table}
\spacingset{2}

\newpage

\newpage

\renewcommand{\thesection}{S\arabic{section}}
\renewcommand{\thefigure}{S\arabic{figure}}
\renewcommand{\thetable}{S\arabic{table}}
\setcounter{section}{0}
\setcounter{figure}{0}
\setcounter{table}{0}

\thispagestyle{empty}

{\bf \Large Supplementary Information: \newline Analyzing recreational fishing effort - Gender differences and the impact of Covid-19}

\newpage
\section{SI Methods}
Of the responses to gender \replaced{12 respondents did not specify their gender and four respondents chose ``Other".}{} 
\replaced{Under}{under} ``Other", one was classified as ``male", two were classified as ``Prefer not to say", and one indicated a nonbinary gender. 
 
This was not included in the quantitative analysis due to the low sample size in the category.

The 11 values in the variables ``Province of residence" and ``Main fishing province" were grouped into five categories for learning Bayesian networks. 
Three categories included one province, respectively: British Columbia (434 and 464 samples), Ontario (431 and 451 samples) and Alberta (491 and 457 samples). 
The fourth category comprised the provinces of Saskatchewan, Manitoba and The Territories (115 and 141 samples) and the fifth category consisted of samples from the provinces of Quebec, New Brunswick, Nova Scotia, Newfoundland and Labrador, and Prince Edward Island (81 and 88 samples). 
Groupings were based on the spatial proximity of the provinces and similar sample sizes.

\newpage
\section{SI Results}
The Angler's Atlas online platform was helpful for the majority to obtain information about maps (76\%) and fish species (62\%) in water bodies, but was used less for information about fishing regulations (33\%), fisher posts (33\%), fishing events (14\%) or as a fishing logbook (12\%). 
The trip report rate, the use of the platform for information on present fish species and on regulations differed between genders (\(\chi^2\) p < 0.01).
More women compared to men reported using the platform to obtain information about fishing regulations (51\% vs. 31\%) and fish species present (74\% vs. 60\%) and reported their trips on the platform (Table \ref{tab:chi-square-results}).

\newpage
\section{SI Figures}
\begin{figure}[H]
    \includegraphics[width=1\linewidth]{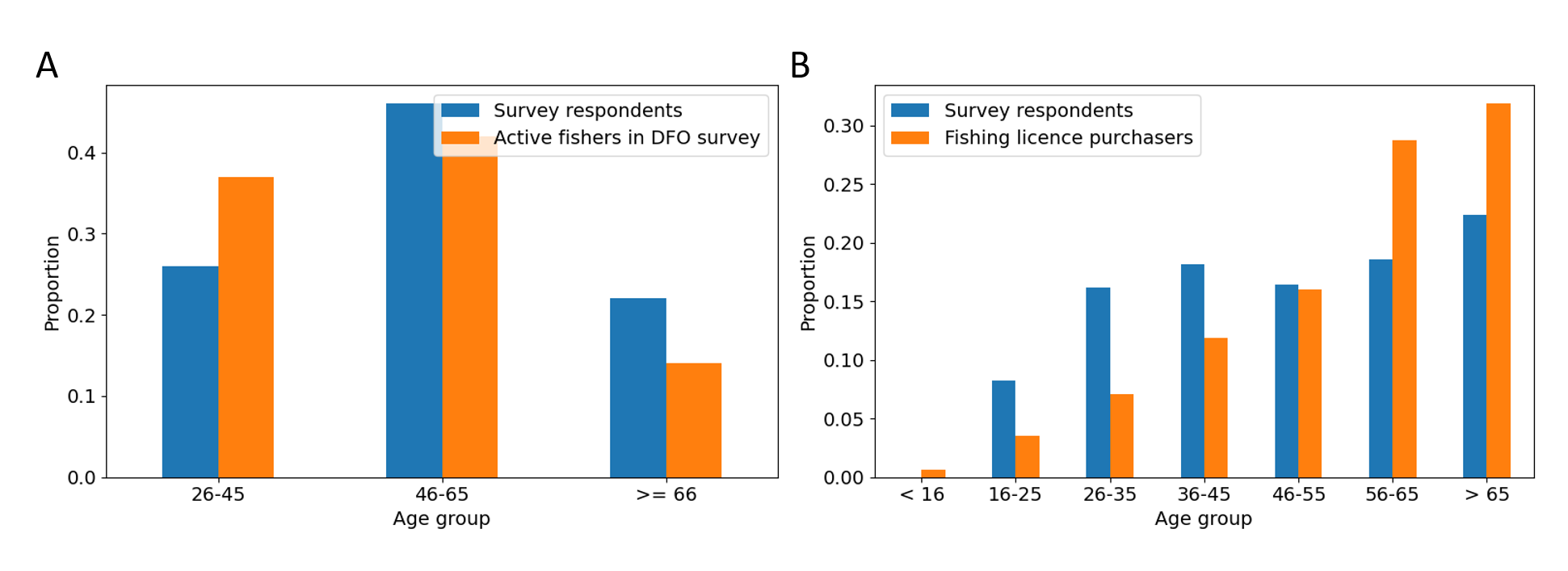}
    \caption{Ages of survey respondents (year 2023) and (A) active fishers in the DFO survey (year 2015) and (B) fishing licence purchasers (year 2022) in the province of British Columbia. Please note that in (A), age ranges of the survey respondents are shown on the x-axis, and age ranges of the DFO survey differed by one year, respectively (e.g., 25-44 years instead of 26-45 years).}
    \label{fig:Age_Distribution}
\end{figure}
  
 \begin{figure}[H]
    \includegraphics[width=1\linewidth]{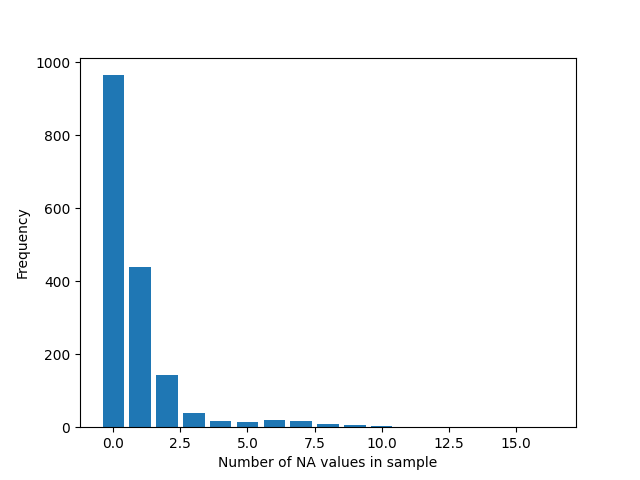}
    \caption{Frequencies of NA values in samples of the entire data set.}
    \label{fig:NAN_values}
\end{figure}
  
\begin{figure}
    \includegraphics[width=1\linewidth]{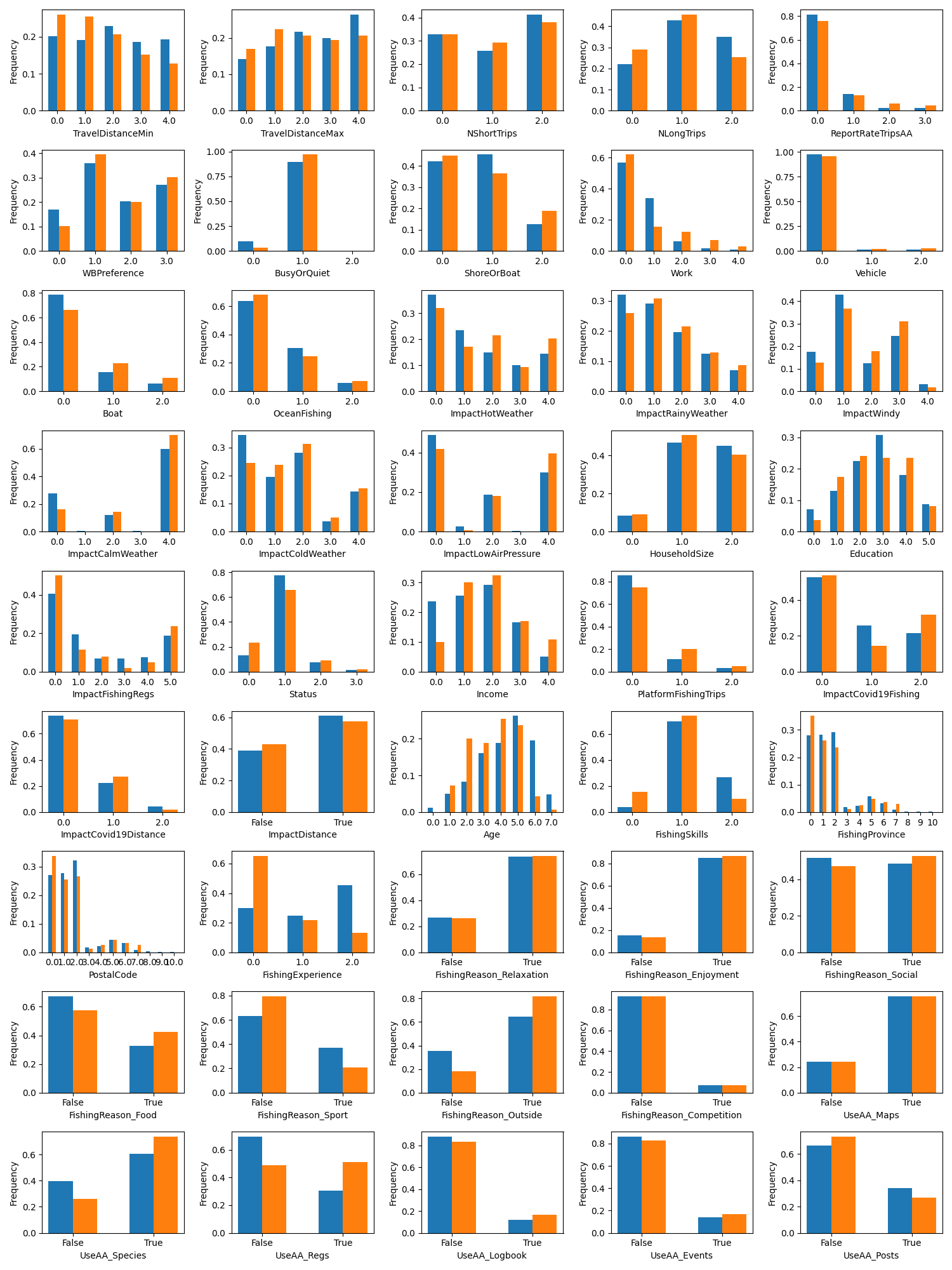}
    \caption{Responses of \replaced{men}{male} (blue) and \replaced{women}{female} (orange)\deleted{ active fishers}. \(\chi^2\) test for difference in proportions of \replaced{women and men}{males and females}, *: p < 0.05, **: p < 0.01. For the meanings of the numbers see Table \ref{tab:Covariates}.} 
    \label{fig:Covariate_Distribution_appendix}
\end{figure}
    
\begin{figure}
    \includegraphics[width=0.63\linewidth]{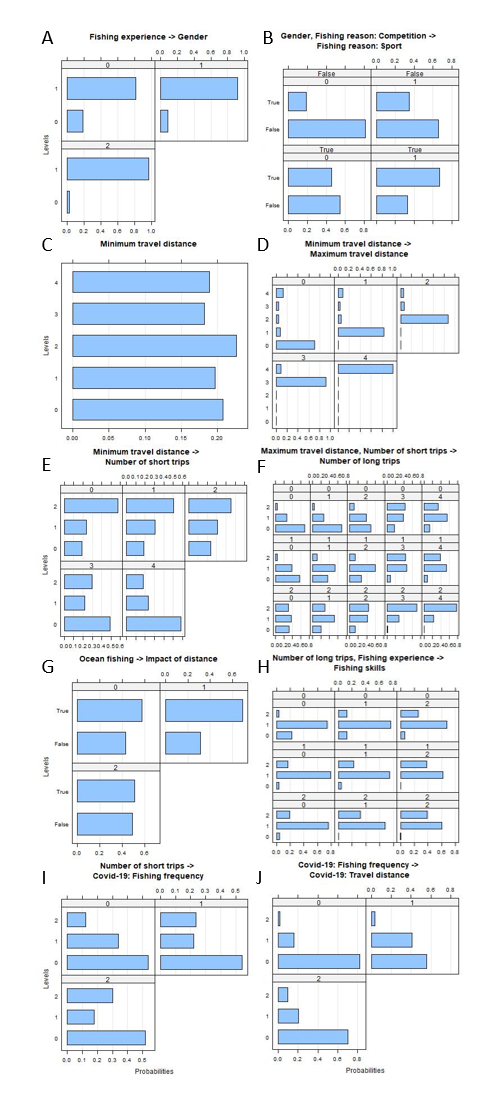}
    \caption{Relationships between variables connected to \replaced{fishing}{fisher} effort. Bars show conditional probabilities.
    Numbers refer to the categories shown in Table \ref{tab:Covariates}.}
    \label{fig:CPDs_appendix}
\end{figure}

\section{SI Tables}

\spacingset{1.2}
\begin{table}[H]
    \centering
    \scriptsize
    \caption{Chi-square independence test results between genders. Significance thresholds: **0.01, *0.05}
    \label{tab:chi-square-results}
    \begin{tabular}{@{}lllll@{}}
        \toprule 
        Category & Variable                   & Chi-Squared & Degrees of freedom & p-value\\ 
        \midrule 
        Demographics &        Age                        & 54.359     & 7.0 & 2.00e-09**                              \\
        and &        Marital status                     & 13.054     & 3.0 & 4.52e-03*                               \\
        socioeconomic status &  Education                  & 9.597      & 5.0 & 8.75e-02                                \\
         & Income                     & 17.419     & 4.0 & 1.60e-03**                               \\
         &  Household size              & 1.172      & 2.0 & 5.57e-01                                \\
         & Work                       & 45.238     & 4.0 & 3.55e-09**                              \\
         & Province of residence                 & 8.796      & 10.0 & 5.52e-01                     \\
         & Vehicle                    & 1.569      & 2.0 & 4.56e-01                                \\
         & Access to boat                       & 13.304     & 2.0 & 1.29e-03**                      \\
        \midrule 
         Fishing &  Fishing experience          & 83.269     & 2.0 & 8.28e-19**                              \\
         characteristics & Fishing skills              & 56.780     & 2.0 & 4.68e-13**                              \\
         & Main fishing province            & 11.229     & 10.0 & 3.40e-01                          \\
         & Impact of distance                           & 0.694      & 1.0 & 4.05e-01          \\
         & Fishing reason: Relaxation  & 0.001      & 1.0 & 9.78e-01                                 \\
         & Fishing reason: Enjoyment   & 0.289      & 1.0 & 5.91e-01                                \\
         & Fishing reason: Social interaction     & 0.947      & 1.0 & 3.31e-01                              \\
         & Fishing reason: Food        & 5.881      & 1.0 & 1.53e-02*                               \\
         & Fishing reason: Sport       & 16.549     & 1.0 & 4.74e-05**                             \\
         & Fishing reason: Being outside     & 19.137     & 1.0 & 1.22e-05**                             \\
         & Fishing reason: Competition & 0.000      & 1.0 & 1.00e+00                              \\
        \midrule 
        Fisher behavior & Frequency short trips                & 1.056      & 2.0 & 5.90e-01        \\ 
         & Frequency long trips                 & 7.623      & 2.0 & 2.21e-02*                     \\
         & Minimum travel distance          & 10.138     & 4.0 & 3.82e-02*                        \\
         & Maximum travel distance          & 4.742      & 4.0 & 3.15e-01                         \\
        \midrule 
        Preferences  & Water-body type             & 5.064      & 3.0 & 1.67e-01         \\ 
        environment & Busy or quiet                     & 8.989      & 2.0 & 1.12e-02*        \\
         & Shore or boat                                & 7.169      & 2.0 & 2.78e-02*        \\
         & Impact fishing regulations                & 17.757     & 5.0 & 3.27e-03**    \\
         & Ocean fishing                                & 2.864      & 2.0 & 2.39e-01      \\
        \midrule 
        Impact weather       & Hot weather           & 10.887     & 4.0 & 2.79e-02*       \\
         & Rainy weather                      & 2.743      & 4.0 & 6.02e-01      \\
         & Windy weather                      & 9.938      & 4.0 & 4.15e-02*       \\
         & Calm weather                       & 11.764     & 4.0 & 1.92e-02*       \\
         & Cold weather                       & 7.038      & 4.0 & 1.34e-01      \\
         & Low air pressure                   & 8.378      & 4.0 & 7.87e-02      \\
        \midrule 
        Usage Angler's  & Use AA: Maps    & 0.000      & 1.0 & 1.00e+00      \\
        Atlas (AA)          & Use AA: Species & 10.524     & 1.0 & 1.18e-03**     \\
         &   Use AA: Regulations         & 27.772     & 1.0 & 1.36e-07**    \\       
         &   Use AA: Logbook             & 2.306      & 1.0 & 1.29e-01     \\
         &   Use AA: Events              & 1.124      & 1.0 & 2.89e-01     \\
         &   Use AA: Posts               & 2.996      & 1.0 & 8.34e-02     \\
         & Platform for fishing trips       & 1.328      & 1.0 & 2.49e-01                 \\
         & Report rate on AA     & 10.386     & 1.0 & 1.56e-02*        \\
        \midrule   
        Impact Covid-19 & Impact Covid-19 on trip frequency & 13.987     & 2.0 & 9.18e-04** \\
                 & Impact Covid-19 on travel distance   & 3.749      & 2.0 & 1.53e-01   \\
        \bottomrule 
    \end{tabular}
\end{table}



\section*{Supplementary material}

1. How far do you typically travel to go fishing?

   *Check all that apply.*

   ☐ Less than 20 km
   ☐ 20 to 50 km
   ☐ 50 to 100 km
   ☐ 100 to 200 km
   ☐ More than 200 km

2. Over the past five years, how many short fishing trips did you go on each year?
   *(i.e., less than 100 km)*

   _____

3. Over the past five years, how many long fishing trips did you go on each year?
   *(i.e., more than 100 km)*

   _____

4. Does the distance you have to travel influence your choice of the water body? And if so, why?

   _____

5. What are your primary reasons you go fishing? (Please select a maximum of three answers)

*Check all that apply.*

- [ ] Relaxation
- [ ] Enjoyment
- [ ] Food
- [ ] Sport
- [ ] Competition (e.g., tournaments)
- [ ] Social experience (family / friends)
- [ ] Being outside
- [ ] Other: _____

6. What kind of water body do you prefer for freshwater fishing?

*Mark only one oval.*

- ( ) Big lake (i.e., tbd: specify surface area here)
- ( ) Small lake (i.e., tbd: specify surface area here)
- ( ) River
- ( ) Not important

7. Why do you prefer this kind of water body?

_____

8. Do you prefer busy or quiet places to fish?

   *Mark only one oval.*

   ◯ Busy places

   ◯ Quiet places

   ◯ Not important

9. Do you usually fish from the shore or a boat?

   *Mark only one oval.*

   ◯ Shore

   ◯ Boat

   ◯ I do both.

10. Why do you prefer to fish from the shore or the boat?

    _____

11. Do fishing regulations (like fish size limits, bag size limits) influence your choice of water body?

    *Mark only one oval.*

    ◯ I don't care about regulations

    ◯ I prefer water bodies with restrictions in fish size and bag size

    ◯ I prefer water bodies with restrictions in fish size

    ◯ I prefer water bodies with restrictions in bag size

    ◯ I prefer water bodies without any restrictions

    ◯ Other: _____

12. Is there anything else that influences the choice of the water body you fish?

    _____

13. Do you also go fishing in the ocean?

    *Mark only one oval.*

    ◯ Yes, I mainly fish in the ocean.
    ◯ Yes, but I prefer freshwater fishing.
    ◯ No.

14. How will these factors influence your fishing?

    *Mark only one oval per row.*

    |  | Usually cancel fishing | Might cancel fishing | Doesn't matter | Might go fishing | Usually going fishing |
    |---|---|---|---|---|---|
    | **Hot weather** | ◯ | ◯ | ◯ | ◯ | ◯ |
    | **Rainy weather** | ◯ | ◯ | ◯ | ◯ | ◯ |
    | **Windy** | ◯ | ◯ | ◯ | ◯ | ◯ |
    | **Calm weather** | ◯ | ◯ | ◯ | ◯ | ◯ |
    | **Cold weather** | ◯ | ◯ | ◯ | ◯ | ◯ |
    | **Low Air Pressure (Cloudy)** | ◯ | ◯ | ◯ | ◯ | ◯ |

15. Is there anything else about the weather that would affect your decision to go fishing or cancel a fishing trip?

    _____

16. What do you find Angler's Atlas helpful for?

    *Check all that apply.*

    ☐ Maps on waterbodies
    ☐ Fish species in waterbodies
    ☐ Fishing regulations
    ☐ MyCatch fishing log book
    ☐ MyCatch fishing events
    ☐ Angler posts
    ☐ Other: _____

17. As a member of Angler's Atlas, what is the most useful information you receive?

    _____

18. Where do you usually record your fishing trip?

    *Mark only one oval.*

    ○ MyCatch App
    ○ Angler's Atlas website
    ○ None of these
    ○ Other: _____

19. Do you record every fishing trip on the MyCatch app or the Angler's Atlas website?

   *Mark only one oval.*

   ◯ Yes

   ◯ More than 50%

   ◯ Less than 50%

   ◯ No trips at all

   ◯ Only successful trips (= trips with catches)

20. Did Covid-19 affect your fishing activity?

   *Mark only one oval.*

   ◯ I went fishing more often

   ◯ Didn't change

   ◯ I went fishing less often

   ◯ Other: _____

21. Did Covid-19 affect the distances you traveled to go fishing?

   *Mark only one oval.*

   ◯ I went fishing closer to home

   ◯ Didn't change

   ◯ I went fishing further from home

   ◯ Other: _____

22. Is there anything else you would like to share with us regarding your ability to go fishing, or things that prevent you from fishing more and should change?

_____
_____
_____
_____
_____

**PART 2: TELL US ABOUT YOURSELF**

This section will help us understand how fishing fits into your broader life.

23. How old are you?

    *Mark only one oval.*

    ◯ Under 16
    ◯ 16 to 25
    ◯ 26 to 35
    ◯ 36 to 45
    ◯ 46-55
    ◯ 56-65
    ◯ 66-75
    ◯ Over 75

24. What is your gender?

    *Mark only one oval.*

    ◯ Male

    ◯ Female

    ◯ Prefer not to say

    ◯ Other: _____

25. What is your marital status?

    *Mark only one oval.*

    ◯ Never married

    ◯ Married

    ◯ Divorced / separated

    ◯ Widowed

    ◯ Other: _____

26. What is your total combined family income for the past 12 months?

    *Mark only one oval.*

    ◯ Under $30,000

    ◯ $30,000 - $60,000

    ◯ $60,000 to $100,000

    ◯ $100,000 to $150,000

    ◯ Over $150,000

    ◯ Choose not answer

27. What is the highest level of education you have completed?

    *Mark only one oval.*

    ◯ 12th grade or less
    ◯ High school graduate
    ◯ Some college / Technical training
    ◯ College graduate
    ◯ University graduate
    ◯ Post graduate degree (Masters or Doctorate)

28. How many people live in your household, including yourself?

    _____

29. What is your employment status?

    *Mark only one oval.*

    ◯ Employed full time
    ◯ Employed part time
    ◯ Unemployed
    ◯ Retired
    ◯ Other: _____

30. Do you own or have easy access to a vehicle?

    *Mark only one oval.*

    ◯ Yes
    ◯ No
    ◯ Sometimes

31. Do you own or have easy access to a boat?

    *Mark only one oval.*

    ◯ Yes

    ◯ No

    ◯ Sometimes

32. How many years have you been fishing?

    _____

33. How would you classify your fishing skills?

    *Mark only one oval.*

    ◯ Beginner

    ◯ Intermediate

    ◯ Expert

34. In which province do you fish the most?

    *Mark only one oval.*

    ◯ Alberta
    ◯ British Columbia
    ◯ Manitoba
    ◯ New Brunswick
    ◯ Newfoundland and Labrador
    ◯ Nova Scotia
    ◯ Ontario
    ◯ Prince Edward Island
    ◯ Quebec
    ◯ Saskatchewan
    ◯ The Territories
    ◯ Other: _____

35. What is your postal code?

    _____

36. Is there anything else you can share that would help us understand the factors the influence your fishing activity?

    _____

This content is neither created nor endorsed by Google.

Google Forms



\end{document}